\documentclass[prd,showpacs,showkeys,nofootinbib,floatfix,
               fleqn,preprint,12pt,tightenlines]{revtex4-1} 

\usepackage{amsmath,amssymb,revsymb,graphicx,dcolumn}

\newcommand{\version}{v3} 

\newcommand{\beq}{\begin{equation}}
\newcommand{\eeq}{\end{equation}}
\newcommand{\beqa}{\begin{eqnarray}}
\newcommand{\eeqa}{\end{eqnarray}}
\newcommand{\bsubeqs}{\begin{subequations}}
\newcommand{\esubeqs}{\end{subequations}}

\newcommand{\ttmp}{t}  
\newcommand{\tautmp}{\tau}  

\newcommand{\sgn}{ \mathrm{sgn}}                

\begin{document}

\begin{widetext}
%
\noindent arXiv:1909.05816 \hfill KA--TP--18--2019\;(\version)
%
\vspace*{3mm}
\end{widetext}

\title{Model for a time-symmetric Milne-like universe without
big bang curvature singularity}%
\vspace*{4mm}

\author{F.R. Klinkhamer}
\email{frans.klinkhamer@kit.edu}
\affiliation{Institute for Theoretical Physics,
Karlsruhe Institute of Technology (KIT),\\
76128 Karlsruhe, Germany}

\author{E. Ling}
\email{eling@math.miami.edu}   
\affiliation{Department of Mathematics,\\
KTH Royal Institute of Technology,
SE-100 44 Stockholm, Sweden\\}

\begin{abstract}
\vspace*{1mm}\noindent
We present a physics model for a time-symmetric Milne-like universe.
The model is
based on the $q$-theory approach to the cosmological constant problem,
supplemented by an assumed vacuum-matter energy exchange
possibly due to quantum-dissipative effects.
Without fine-tuning of the initial vacuum energy density,
we obtain a harmless big bang singularity (with finite values of the
Ricci and Kretschmann curvature scalars)
and attractor behavior towards Minkowski spacetime.
The time-symmetric spacetime manifold of our model may provide the
proper setting for a possible CPT-symmetric universe.  
\\
\vspace*{-0mm}
\end{abstract}

\pacs{04.20.Cv, 98.80.Bp, 98.80.Jk}
\keywords{general relativity, big bang theory,
          mathematical and relativistic aspects of cosmology}

\maketitle

\section{Introduction}
\label{sec:Intro}

Certain singular versions of the Friedmann 
equation~\cite{Friedmann1922-1924,MisnerThorneWheeler2017}
have recently been shown to have nonsingular solutions, i.e., solutions
without physical big bang
singularities~\cite{Ling2018,Klinkhamer2019,Klinkhamer2019-revisited}.

Here, we revisit the Milne-like universe of Ref.~\cite{Ling2018},
where the spacetime near the apparent big bang singularity
resembles the spacetime of the 
so-called Milne universe~\cite{Milne1932}
(see also Sec.~5.3 of Ref.~\cite{BirrellDavies1982}
and Sec.~1.3.5 of Ref.~\cite{Mukhanov2005}),
which is just a slice of Minkowski spacetime in expanding coordinates
and, hence, has no physical singularity.
In this way, we are able to improve upon the metric
used in Ref.~\cite{BoyleFinnTurok2018}
for a ``CPT-symmetric universe,''
as that metric does have a physical big bang singularity
with, for example, a diverging Kretschmann curvature scalar.

The goal, now, is to construct an explicit physics model
for a time-symmetric Milne-like universe.
For the particular model, we will use $q$-theory
(see Refs.~\cite{KlinkhamerVolovik2008a,KlinkhamerVolovik2008b,%
KlinkhamerVolovik2016}
and references therein), because it is, then, possible to describe
the change of vacuum energy in a natural and consistent way.

Specifically, the model has two characteristics. First, there is,
in the simplest version of the model, only an arbitrary
positive cosmological constant operative at $\ttmp=0$,
which transforms the big bang singularity of an appropriate
spacetime metric into a coordinate singularity~\cite{Ling2018}.
Second, there is the $q$-field which ultimately compensates
the cosmological constant,
so that Minkowski spacetime is approached at late times.
The focus of our model is on 
rendering the big bang singularity at $\ttmp=0$ harmless
and not on getting an entirely realistic description
at later times.

\section{Model for a Milne-like universe}
\label{sec:Model-for-a-Milne-like-universe}

\subsection{Brief description}  
\label{subsec:Brief-description}

The metric obtained in this article improves upon the
one used in Ref.~\cite{BoyleFinnTurok2018}, as it does
not have a big bang curvature singularity~\cite{Ling2018}:
the spacetime near the apparent big bang singularity
resembles the spacetime of the Milne
universe~\cite{Milne1932,BirrellDavies1982,Mukhanov2005},
which corresponds to a slice of Minkowski spacetime
in expanding coordinates
and, therefore, has no curvature singularity.
An alternative way to control the big bang singularity
has been presented in
Refs.~\cite{Klinkhamer2019,Klinkhamer2019-revisited},
but will not be discussed further here.

We now start with the construction of our physics model for
a Milne-like universe.
In the framework of $q$-theory~\cite{KlinkhamerVolovik2008a},
we use the extended
model of Ref.~\cite{KlinkhamerVolovik2016} with minor changes.
The main inputs are as follows:
\begin{itemize}
  \item
a spatially-hyperbolic Robertson--Walker metric,
  \item
a conserved relativistic $q$-field,
  \item
a relativistic matter component,  
  \item
special vacuum-matter energy exchange,
  \item
special boundary conditions at $\ttmp=0$, which give
a positive vacuum energy density, $\rho_V(0)>0$, and a
vanishing matter energy density, $\rho_{M}(0)=0$.
\end{itemize}

The outline of the rest of this section is as follows.
The metric will be discussed in Sec.~\ref{subsec:Metric}
and the details of the theory in Sec.~\ref{subsec:Theory},
together with some appropriate \textit{Ans\"{a}tze}.
The reduced field equations are obtained in
Sec.~\ref{subsec:ODEs} and analytic and numerical
results are presented in Secs.~\ref{subsec:Analytic-results}
and \ref{subsec:Numerical-results}, respectively.
Finally, we mention some generalizations in
Sec.~\ref{subsec:Generalizations} and discuss the
attractor behavior towards Minkowski spacetime.

\subsection{Metric}
\label{subsec:Metric}

The metric \textit{Ansatz} is given by
the spatially-hyperbolic ($k=-1$)  Robertson--Walker 
metric~\cite{MisnerThorneWheeler2017}  
in terms of comoving spatial 
coordinates $\{\chi,\, \vartheta,\, \varphi\}$, 
\bsubeqs\label{eq:RW-metric-k-is-minus-1-ranges-coordinates}
\beqa\label{eq:RW-metric-k-is-minus-1}
\hspace*{-2mm}
ds^{2}
&\equiv&
g_{\mu\nu}(x)\, dx^\mu\,dx^\nu
=
- d\ttmp^{2}
+ a^{\,2}(\ttmp)\,\left[d\chi^2+\sinh^2(\chi) \,
\Big( d\vartheta^2+\sin^2(\vartheta)\, d\varphi^2 \Big)\right]\,,
\\[2mm]
\label{eq:ranges-coordinates}
\ttmp &\in& (-\infty,\,\infty)\,,
\quad
\chi \in [0,\,\infty)\,, 
\quad
\vartheta \in [0,\,\pi]\,, 
\quad
\varphi \in [0,\,2\pi)\,,
\eeqa
\esubeqs
where the infinite range of the cosmic time coordinate $t$  
will be clarified later on.

The resulting expressions for the
Ricci curvature scalar
$R(x) \equiv g^{\nu\sigma}(x)\,g^{\mu\rho}(x)\,R_{\mu\nu\rho\sigma}(x)$
and the Kretschmann curvature scalar
$K(x) \equiv R^{\mu\nu\rho\sigma}(x)\,R_{\mu\nu\rho\sigma}(x)$ are
\bsubeqs\label{eq:R-K-from-a}
\beqa
\label{eq:R-from-a}
R[a(\ttmp)] &=&
6\,\left[\frac{\ddot{a}(\ttmp)}{a(\ttmp)}
           + \left(\frac{\dot{a}(\ttmp)}{a(\ttmp)}\right)^2
           -\frac{1}{a^2(\ttmp)}\right]\,,
\\[2mm]
\label{eq:K-from-a}
K[a(\ttmp)] &=&
12\,\left[\left(\frac{\ddot{a}(\ttmp)}{a(\ttmp)}\right)^2
           + \left(\left(\frac{\dot{a}(\ttmp)}{a(\ttmp)}\right)^2
           -\frac{1}{a^2(\ttmp)}\right)^2\right] \,,
\eeqa
\esubeqs
where the overdot stands for differentiation with respect to $\ttmp$.
Observe that these curvature scalars vanish
for $a(\ttmp)=\ttmp$, which is precisely the scale factor of the
Milne universe~\cite{Milne1932,BirrellDavies1982,Mukhanov2005}.

\subsection{Theory}
\label{subsec:Theory}

For the $q$-theory considered, we use the four-form-field-strength
realization of the
$q$-field \cite{KlinkhamerVolovik2008a,KlinkhamerVolovik2008b}
and neglect any possible modification of Einstein's general relativity
[which would, for example, have a gravitational coupling parameter
$G=G(q) \ne G_N=\text{constant}$].
The classical action is given by:
\bsubeqs\label{eq:classical-action}
\beqa
&&
I_\text{class.}=
\int_{\mathbb{R}^{4}}
\,d^{4}x\, \sqrt{-g}\,\left(\frac{R}{16\pi\, G_N} +\epsilon(q)
+\mathcal{L}^\text{\,M}(\psi)\right) \,,
\label{eq:actionF}\\[2mm]
&&
q^2 \equiv- \frac{1}{24}\,
F_{\kappa\lambda\mu\nu}\,F^{\kappa\lambda\mu\nu}\,,\quad
F_{\kappa\lambda\mu\nu}\equiv
\nabla_{[\kappa}A_{\lambda\mu\nu]}\,,
\label{eq:Fdefinition1}\\[2mm]
&&
F_{\kappa\lambda\mu\nu}=q\sqrt{-g} \,\epsilon_{\kappa\lambda\mu\nu}\,,\quad
F^{\kappa\lambda\mu\nu}=q \,\epsilon^{\kappa\lambda\mu\nu}/\sqrt{-g}\,,
\label{eq:Fdefinition2}
\eeqa
\esubeqs
where $F$ is the four-form field strength from a three-form
gauge field $A$ (Ref.~\cite{KlinkhamerVolovik2008a} contains an
extensive list of references) and $\epsilon(q)$ is a generic even function
of $q$. The symbol $\nabla_\mu$ in \eqref{eq:Fdefinition1} stands for
the covariant derivative and a pair of square brackets
around spacetime indices denotes complete anti-symmetrization.
The symbol $\epsilon_{\kappa\lambda\mu\nu}$ in \eqref{eq:Fdefinition2}
corresponds to the Levi--Civita symbol, which makes $q$  a pseudoscalar
and $q^2$ a scalar.
Hence, $q^2$ in the action \eqref{eq:actionF} is a scalar
but not a fundamental scalar, $q^2=q^2(A,\,g)$.
The gauge field $A_{\mu\nu\rho}(x)$,
the metric $g_{\mu\nu}(x)$, and the generic matter field $\psi(x)$
are the fundamental fields of the theory.
Here, and in the following, we use natural units with
$c=1$ and $\hbar=1$.

The field equations from \eqref{eq:classical-action}
are the Einstein equation with
the gravitating vacuum energy density $\rho_V(q)\ne \epsilon(q)$ and
the generalized Maxwell equation,
\bsubeqs\label{eq:Einstein-Maxwell}
\begin{eqnarray}
\hspace*{-0mm}
&&
\frac{1}{8\pi G_{N}}
\left( R_{\mu\nu}-\frac{1}{2}\,R\,g_{\mu\nu}\right)
=
\rho_{V}(q)\,g_{\mu\nu}
-T^\text{\,M}_{\mu\nu}\,,
\label{eq:EinsteinEquation2}\\[2mm]
\hspace*{-0mm}
&&
\nabla_\nu \left(\sqrt{-g} \;\frac{F^{\kappa\lambda\mu\nu}}{q}
\;\frac{d\epsilon(q)}{d q}  \right)=0\,,
\label{eq:MaxwellEquation2}
\\[2mm]
\label{eq:rhoV-def}
\hspace*{-0mm}
&&
\rho_{V}(q)  \equiv   \epsilon(q) -q\,\frac{d\epsilon(q)}{d q}  \,,
\end{eqnarray}
\esubeqs
where $T^\text{\,M}_{\mu\nu}$ is the matter energy-momentum tensor
from \eqref{eq:actionF}.

We now turn to the metric \textit{Ansatz} 
\eqref{eq:RW-metric-k-is-minus-1-ranges-coordinates}
and consider a homogenous $q$-field,
\beq\label{eq:homogenous-q}
q=q(t)\,.
\eeq
The matter content is assumed to be given by
a homogenous perfect fluid with the following
energy density, pressure, and equation-of-state parameter,
\bsubeqs\label{eq:rhoM-PM-wM-constant}
\beqa
\rho_{M} &=& \rho_{M}(t)\,,
\\[2mm]
P_{M}&=& P_{M}(t)\,,
\\[2mm]
\label{eq:wM-constant}  
W_{M}(t) &\equiv& P_{M}(t)/\rho_{M}(t) = w_{M} \geq 0\,,
\eeqa
\esubeqs
where $w_{M}$ is a nonnegative constant.
For this setup with  
\textit{Ans\"{a}tze} \eqref{eq:RW-metric-k-is-minus-1-ranges-coordinates},
\eqref{eq:homogenous-q}, and  \eqref{eq:rhoM-PM-wM-constant},
the field equations \eqref{eq:Einstein-Maxwell}
produce reduced field equations
which correspond to ordinary differential equations (ODEs).
These ODEs will be given in Sec.~\ref{subsec:ODEs}
in terms of dimensionless variables.

At this moment, we need to mention one last important
property of the theory considered.
This concerns possible quantum-dissipative effects that
produce vacuum-matter energy exchange~\cite{KlinkhamerVolovik2016}.
In practical terms, these effects are modelled by 
a source term on the right-hand side of the
generalized Maxwell equation \eqref{eq:MaxwellEquation2},
with the corresponding opposite term in the matter energy-conservation
equation. In the cosmological context, we have
\bsubeqs\label{eq:rhoVdoteq-rhoMdoteq}
\beqa
\label{eq:rhoVdoteq}
-q\,\partial_{\ttmp} \left(\frac{d\epsilon(q)}{d q} \right)
=
\partial_{\ttmp}  \rho_{V}(q)
=-\mathcal{S}\,,
\\[2mm]
\label{eq:rhoMdoteq}
\partial_{\ttmp}\,\rho_{M} +3\,H \,(1+w_{M})\,\rho_{M}
=+\mathcal{S}\,,
\eeqa
\esubeqs
with the first equality in \eqref{eq:rhoVdoteq}
resulting from the $\rho_{V}$ definition \eqref{eq:rhoV-def}
and the Hubble parameter in \eqref{eq:rhoMdoteq}
has the standard definition,
\beq\label{eq:H-def}
H(\ttmp) \equiv a^{-1}(\ttmp)\, [d a(\ttmp)/d \ttmp]\,.
\eeq

The explicit calculation~\cite{KlinkhamerVolovik2016}
of the production of massless particles (gravitons)   
by spacetime
curvature~\cite{ZeldovichStarobinsky1977}
gives a particular result for the source term:
$\mathcal{S} \propto \hbar\,H\,R^2$  
with
Hubble parameter $H$ and Ricci curvature scalar $R$.
Here, we assume a somewhat different functional
dependence on the cosmic scale factor $a(\ttmp)$,
\beq\label{eq:S-assumed}
\mathcal{S}(\ttmp) = \Gamma\;
\frac{a(\ttmp)}{\sqrt{a^2(\ttmp)+1}}\;
\Big(R^2(\ttmp)/36\Big)\;l^{4}_\text{decay}\,,
\eeq
with the Ricci curvature scalar $R(\ttmp)$ from \eqref{eq:R-from-a},
a positive decay constant $\Gamma$, and
a length scale $l_\text{decay}$.
In the following, we assume that $l_\text{decay}$
is equal to the Planck length scale,
\beq\label{eq:l-planck}
l_\text{planck} \equiv \hbar c/E_\text{planck}
\equiv \sqrt{8\pi G_N \hbar/c^{3}}
\approx 8.10 \times 10^{-35}\,\text{m}\,,
\eeq
with $c$ and $\hbar$ temporarily displayed
(the numerical value of the reduced Planck energy is  
$E_\text{planck} \approx 2.44 \times 10^{18}\,\text{GeV}$).
%
%

Even though we do not have a detailed derivation of the
source term as shown in \eqref{eq:S-assumed},
it suffices for the purpose of obtaining
an ``existence proof'' of a physics model without
big bang curvature singularity.
Further discussion of the source term
is postponed to Secs.~\ref{subsec:ODEs} and \ref{sec:Discussion}.

\subsection{ODEs}
\label{subsec:ODEs}

We now introduce dimensionless variables ($\hbar=c=8 \pi G_{N} =1$)
as in Sec. IV B of Ref.~\cite{KlinkhamerVolovik2008b}
or in Sec. 6 of Ref.~\cite{KlinkhamerVolovik2016}.
Then, $f(\tautmp)$ is the dimensionless variable corresponding to $q(t)$,
$r_V(\tautmp)$ the dimensionless variable corresponding to $\rho_V(t)$,
and $r_{M}(\tautmp)$ the dimensionless variable corresponding to $\rho_{M}(t)$,
where $\tautmp$ is the dimensionless cosmic time coordinate.
The ODEs for $a(\tautmp)$, $f(\tautmp)$, and $r_{M}(\tautmp)$
are essentially the same as 
those in Ref.~\cite{KlinkhamerVolovik2016}:   
\bsubeqs\label{eq:ODEs-Milne-like}
\beqa
%
&&\frac{4/3}{1+w_{M}}\,\dot{h} +2\,h^2
-\frac{2/3+2\,w_{M}}{1+w_{M}}\,\frac{1}{a^2}
= \frac{2}{3}\,r_{V}\,,
\label{eq:adotdot-ODE-Milne-like}
\\[2mm]
&&h^2-\frac{1}{a^2}
= \frac{1}{3}\,\Big(r_{V}+r_{M} \Big)\,,
\label{eq:adot-ODE-Milne-like}
\\[2mm]
&&f\;\dot{f} \;\epsilon^{\prime\prime}(f)=
\gamma\;\frac{a}{\sqrt{a^2+1}}\;
\left(\frac{2}{3}\,r_{V}\right)^2\,,
\label{eq:fdot-ODE-Milne-like}
\\[2mm]
\label{eq:h-def-Milne-like}
&&h = \dot{a}/a\,,
\\[2mm]
\label{eq:rV-def-Milne-like}
&&r_{V}(f)  =  \epsilon(f) - f\,\epsilon^{\prime}(f)\,,
\\[2mm]
\label{eq:pos-lambda--Milne-like}
&&\gamma  >  0\,,
\eeqa
\esubeqs
where, for the rest of this section, 
the overdot stands for differentiation with respect to
$\tautmp$ and the prime for differentiation with respect to $f$.
According to \eqref{eq:adotdot-ODE-Milne-like},
the source term on the right-hand side   
of \eqref{eq:fdot-ODE-Milne-like}
reproduces, for the case of relativistic matter ($w_{M}=1/3$),
the expression \eqref{eq:S-assumed}
in terms of the Ricci scalar $R=6\,\big(\dot{h} +2\,h^2-1/a^2\big)$
from \eqref{eq:R-from-a}.

As mentioned in Sec.~\ref{subsec:Theory},
the crucial new input of these model equations is the
right-hand side of \eqref{eq:fdot-ODE-Milne-like},
where the $a$ numerator makes for a vanishing $\dot{f}$ if $a$ vanishes
(as happens for the apparent big bang singularity at $t=0$,
see Sec.~\ref{subsec:Analytic-results})
and where the $(2\,r_{V}/3)^2$ term makes for
a decreasing vacuum-matter energy exchange as Minkowski spacetime is
approached [with $r_{V} \to 0$ for $|\tautmp|\to \infty$,
as will be shown in Sec.~\ref{subsec:Analytic-results}].
A further property of \eqref{eq:fdot-ODE-Milne-like}
is its invariance under the following time-reversal transformation:
\beq
\tautmp \to -\tautmp\,, \quad
a(\tautmp) \to -a(-\tautmp)\,, \quad
f(\tautmp) \to f(-\tautmp)\,,
\eeq
where the odd behavior of the scale factor $a(\tautmp)$ is a characteristic  
of the Milne-like universe that is possibly relevant to fermionic fields.

In order to be specific,
we consider a matter component with an
ultrarelativistic equation-of-state parameter,   
\bsubeqs\label{eq:wM-epsilon-lambda-Ansatz-Milne-like}
\beqa\label{eq:wM-Ansatz-Milne-like}
w_{M} &=& 1/3\,,
\eeqa
and choose the following \textit{Ansatz} function:
\beqa
\label{eq:epsilon-Ansatz-Milne-like}
\epsilon(f) &=& \lambda+ f^2 + 1/f^2\,,
\\[2mm]
\label{eq:lambda-Ansatz-Milne-like}
\lambda &>& 0 \,,
\eeqa
\esubeqs
whose properties have been discussed in the sentence
below Eq.~(6.3b) of Ref.~\cite{KlinkhamerVolovik2016}.
[As will be discussed in Sec.~\ref{subsec:Generalizations},
the condition \eqref{eq:lambda-Ansatz-Milne-like} can be relaxed,   
but, for now, we consider only a positive cosmological constant.]
For the \textit{Ansatz} \eqref{eq:epsilon-Ansatz-Milne-like}, the
dimensionless gravitating vacuum energy density \eqref{eq:rV-def-Milne-like}
is given by
\bsubeqs\label{eq:rV-f0-Ansatz-Milne-like}
\beqa\label{eq:rV-Ansatz-Milne-like}
r_{V}&=&\lambda+3/f^2 -f^2\,,
\eeqa
which vanishes for the following equilibrium value
(real and positive):
\beqa\label{eq:f0-Ansatz-Milne-like}
f_0 &=& \sqrt{\Big(\lambda + \sqrt{12 +\lambda^2}\,\Big)\Big/2}\,.
\eeqa
\esubeqs

The ODEs \eqref{eq:adotdot-ODE-Milne-like},
\eqref{eq:adot-ODE-Milne-like},
and \eqref{eq:fdot-ODE-Milne-like}
are to be solved simultaneously.
The boundary conditions at $\tautmp=0$ are as follows:
\bsubeqs\label{eq:a-f-rM-bcs-Milne-like}
\beqa
\label{eq:a-bcs-Milne-like}
a(0) &=&     0\,,
\\[2mm]
\label{eq:f-bcs-Milne-like}
f(0) &=&      f_\text{bb} > 0\,,\quad
f_\text{bb} \ne f_0\,,
\\[2mm]
\label{eq:rM-bcs-Milne-like}
r_{M}(0) &=& 0 \,.
\eeqa
\esubeqs
With boundary condition \eqref{eq:a-bcs-Milne-like},
the Friedmann ODEs \eqref{eq:adotdot-ODE-Milne-like} and
\eqref{eq:adot-ODE-Milne-like} are singular at $\tautmp=0$
due to the terms $k/a^2$ for $k=-1$.
But, by construction, the solution near $\tautmp=0$ is essentially
known~\cite{Ling2018}, as will be shown in Sec.~\ref{subsec:Analytic-results}.

\subsection{Analytic results}
\label{subsec:Analytic-results}

We can now use the same procedure as in Ref.~\cite{Klinkhamer2019-revisited}
for the regularized Friedmann singularity.
Specifically, we obtain, for a small interval
$\tautmp \in [-\Delta\tautmp,\,\Delta\tautmp]$,  an
approximate analytic solution of the ODEs \eqref{eq:ODEs-Milne-like}
with $w_{M} = 1/3$ and, for $\tautmp$ outside this
interval, the numerical solution with
matching boundary conditions at $\tautmp =\pm \Delta\tautmp$.
Analytic results are discussed in this subsection
and numerical results in Sec.~\ref{subsec:Numerical-results}.

The zeroth-order analytic solution is given by
\bsubeqs\label{eq:zeroth-order-analytic-solution-Milne-like}
\beqa
\label{eq:zeroth-order-analytic-a-solution-Milne-like}
\overline{a}(\tautmp) &=& \frac{1}{\sqrt{\lambda/3}}\;
\sinh\left(\sqrt{\lambda/3}\;\tautmp\right) \,,
\\[2mm]
\label{eq:zeroth-order-analytic-f-solution-Milne-like}
\overline{f}(\tautmp)  &=& \sqrt[4]{3} \approx 1.3161\,,  
\\[2mm]
\label{eq:zeroth-order-analytic-rM-solution-Milne-like}
\overline{r}_{M}(\tautmp)   &=& 0\,,
\eeqa
\esubeqs
where the particular $f$ value
in \eqref{eq:zeroth-order-analytic-f-solution-Milne-like} makes that
the corresponding value of the gravitating vacuum energy density
\eqref{eq:rV-Ansatz-Milne-like}
is solely given by the cosmological constant, $\overline{r}_V(\tautmp)=\lambda$. Observe that,
for $\tautmp\to 0$ and any fixed positive value of $\lambda$,
the scale factor function
\eqref{eq:zeroth-order-analytic-a-solution-Milne-like}
approaches the function $a_\text{Milne}(\tautmp)=\tautmp$ of
the Milne
universe~\cite{Milne1932,BirrellDavies1982,Mukhanov2005},
which solves the ODEs
\eqref{eq:adotdot-ODE-Milne-like}, \eqref{eq:adot-ODE-Milne-like}, and
\eqref{eq:h-def-Milne-like}
for $r_V=r_{M}=0$.

Two remarks on the zeroth-order analytic solution
\eqref{eq:zeroth-order-analytic-solution-Milne-like}
are in order. First, we get the following expressions
for the Ricci curvature scalar $R$
and the Kretschmann curvature scalar
$K$ from \eqref{eq:R-K-from-a}:
\bsubeqs\label{eq:R-K-from-abar}
\beqa
R[\overline{a}(\tautmp)] &=& 4\,\lambda\,,
\\[2mm]
K[\overline{a}(\tautmp)] &=& \frac{8}{3}\;\lambda^2\,,
\eeqa
\esubeqs
which are constant (i.e., $\tautmp$-independent), 
as we really have a slice  
of de-Sitter spacetime~\cite{Ling2018}.
Second, the functions \eqref{eq:zeroth-order-analytic-solution-Milne-like}
provide, for $\gamma=0$, the exact solution of the ODEs \eqref{eq:ODEs-Milne-like}
over the whole cosmic time range ($\tautmp \in \mathbb{R}$)
and, for this reason, we have
called \eqref{eq:zeroth-order-analytic-solution-Milne-like}
the zeroth-order analytic solution.


We can push the analytic calculation somewhat further.
Consider the following perturbative \textit{Ans\"{a}tze}:%
\bsubeqs\label{eq:pert-solution-Milne-like}
\beqa
\label{eq:pert-a-solution-Milne-like}
a_\text{pert}(\tautmp) &=&
\tautmp + a_{3}\,\tautmp^{3}+ a_{5}\,\tautmp^{5}+ a_{7}\,\tautmp^{7}
+ \ldots \,,
\\[2mm]
\label{eq:pert--f-solution-Milne-like}
f_\text{pert}(\tautmp)  &=& 3^{1/4}\,
\Big[ 1+ c_{2}\,\tautmp^{2}+ + c_{4}\,\tautmp^{4}+ c_{6}\,\tautmp^{6}
+ \ldots \Big]\,.
\eeqa
\esubeqs
Inserting these \textit{Ans\"{a}tze}
into the ODEs \eqref{eq:adotdot-ODE-Milne-like}
and \eqref{eq:fdot-ODE-Milne-like}
gives the following results for the first few coefficients:
\bsubeqs\label{eq:coeff}
\beqa
a_3 &=& \frac{\lambda}{18}\,,
\\[2mm]
a_5 &=& \frac{\lambda^2}{3240}\;\Big[ 3 - 16\,\gamma \Big]\,,
\\[2mm]
a_7 &=& \frac{\lambda^2}{408240}\;
\Big[ 3\, \lambda + 270\,\gamma  - 86\, \gamma\, \lambda
      + 240\, \gamma^2\, \lambda\, \Big]\,,
\\[2mm]
c_2 &=&
\gamma\;\frac{\lambda^2}{18\,\sqrt{3}}\,,
\\[2mm]
c_4 &=& \gamma\;\frac{\lambda^2}{1944\,{\sqrt{3}}}\;
        \left[ -27 + \left(
        3 - 24\,\gamma \right) \,\lambda + {\sqrt{3}}\,\gamma\,\lambda^2\,
        \right]\,,
\\[2mm]
c_{6} &=&  c_{61}\,\gamma+c_{62}\,\gamma^2+c_{63}\,\gamma^3\,,
\eeqa
\esubeqs
where the coefficients $c_{61}$, $c_{62}$, and $c_{63}$
are rather bulky and are not given explicitly here.

With $a_\text{pert}(\tautmp)$ and $f_\text{pert}(\tautmp)$ in hand,
we obtain  $r_\text{V,\,pert}(\tautmp)$
from \eqref{eq:rV-Ansatz-Milne-like}
and $r_\text{M,\,pert}(\tautmp)$ from \eqref{eq:adot-ODE-Milne-like},
\bsubeqs\label{eq:pert-solution-rV-and-rM}
\beqa\label{eq:pert-solution-rV}
r_{V,\,\text{pert}}(\tautmp)
&=&
\lambda - \gamma\;\frac{2}{9}\,\lambda^2\,\tautmp^2
+\gamma\;\frac{1}{162}
\,\lambda^2\,\Big[ 9 - \left( 1 - 8\,\gamma \right) \,\lambda\,\Big]\tautmp^{4}
+  \text{O}(\tautmp^6)\,,
\\[2mm]
\label{eq:pert-solution-rM}
r_{M,\,\text{pert}}(\tautmp)
&=&
\gamma\;\frac{2}{27}\,\lambda^2\tautmp^2
- \gamma\;\frac{1}{972}\,\lambda^2 \;
\Big[27 + \lambda + 24\, \gamma\, \lambda\,\Big]\,\tautmp^{4}
+  \text{O}(\tautmp^6)\,.
\eeqa
\esubeqs
Considering the $\text{O}(\tautmp^2)$ terms
in \eqref{eq:pert-solution-rV-and-rM}, we note that
energy transferred from the vacuum to the matter component
does two things thermodynamically:
it increases the matter energy density
and it lets the extra matter perform work.
With $a_\text{pert}(\tautmp)$, we obtain
the following series expansions of the dimensionless Ricci curvature scalar $R$
and the dimensionless Kretschmann curvature scalar $K$
from \eqref{eq:R-K-from-a}:
\bsubeqs\label{eq:pert-solution-R-and-K}
\beqa
R_\text{pert}(\tautmp) &=&
4\,\lambda
-
\frac{8}{9}\; \gamma\; \lambda^2\,\tautmp^2
+ \frac{2}{81}\;\gamma\;\lambda^2
\big[ 9 - \left( 1 - 8\, \gamma \right) \, \lambda \big]\,\tautmp^{4}
+  \text{O}(\tautmp^6) \,,
\\[2mm]
K_\text{pert}(\tautmp) &=&
\frac{8}{3}\;\lambda^2
-\frac{32}{27}\;\gamma\, \lambda^3\,\tautmp^2
+
\frac{8}{2187}\; \gamma\, \lambda^3\;
\big[ 81 - \left( 9 - 112\, \gamma \right) \, \lambda \big]
\,\tautmp^{4}
+  \text{O}(\tautmp^6) \,,
\eeqa
\esubeqs
which show that $R(\tautmp)$ and $K(\tautmp)$ are nonsingular 
and have a local maximum at $\tautmp=0$.

The ODEs \eqref{eq:ODEs-Milne-like} are relatively simple and we
obtain the following asymptotic results for $|\tautmp|\to \infty$:
\bsubeqs\label{eq:asymptotic-results}
\beqa
\label{eq:asymptotic-h}
h
&\sim&
\frac{1}{2}\,\sqrt{3/\gamma}\;\text{sgn}(\tautmp)\;\big(\tautmp^2\big)^{-1/4}\,,
\\[2mm]
\label{eq:asymptotic-rV}
r_V &\sim& 9/(4\gamma)\;\big(\tautmp^2\big)^{-1/2}\,,
\\[2mm]
\label{eq:asymptotic-rM-over-rV}
r_{M}/r_V &\to& 0\,,
\eeqa
\esubeqs
with the sign function
\begin{eqnarray}
\label{eq:sgn-def}
\hspace*{-0mm}
\text{sgn}(x)&\equiv&
\begin{cases}
 x/\sqrt{x^{2}}\,,    & \;\;\text{for}\;\; x\ne 0\,,
 \\[2mm]
 0\,,    & \;\;\text{for}\;\; x=0\,.
\end{cases}
\end{eqnarray}
From \eqref{eq:asymptotic-h} without further correction terms,
we have an inflation-type expansion,
\beqa\label{eq:inflation-type-expansion}
a(\tautmp) &\propto& \text{sgn}(\tautmp)\;
\exp\Big[\sqrt{3/\gamma}\;\big(\tautmp^2\big)^{1/4}\Big],
\eeqa
similar to what was found in Ref.~\cite{KlinkhamerVolovik2016}.
Incidentally,   
the asymptotic results \eqref{eq:asymptotic-results} and
\eqref{eq:inflation-type-expansion} hold not only
for $w_{M}=1/3$ but for any value of $w_{M}$ larger than $-1$.



\begin{figure*}[t] 
\vspace*{-0mm}
\begin{center}
\includegraphics[width=1\textwidth]{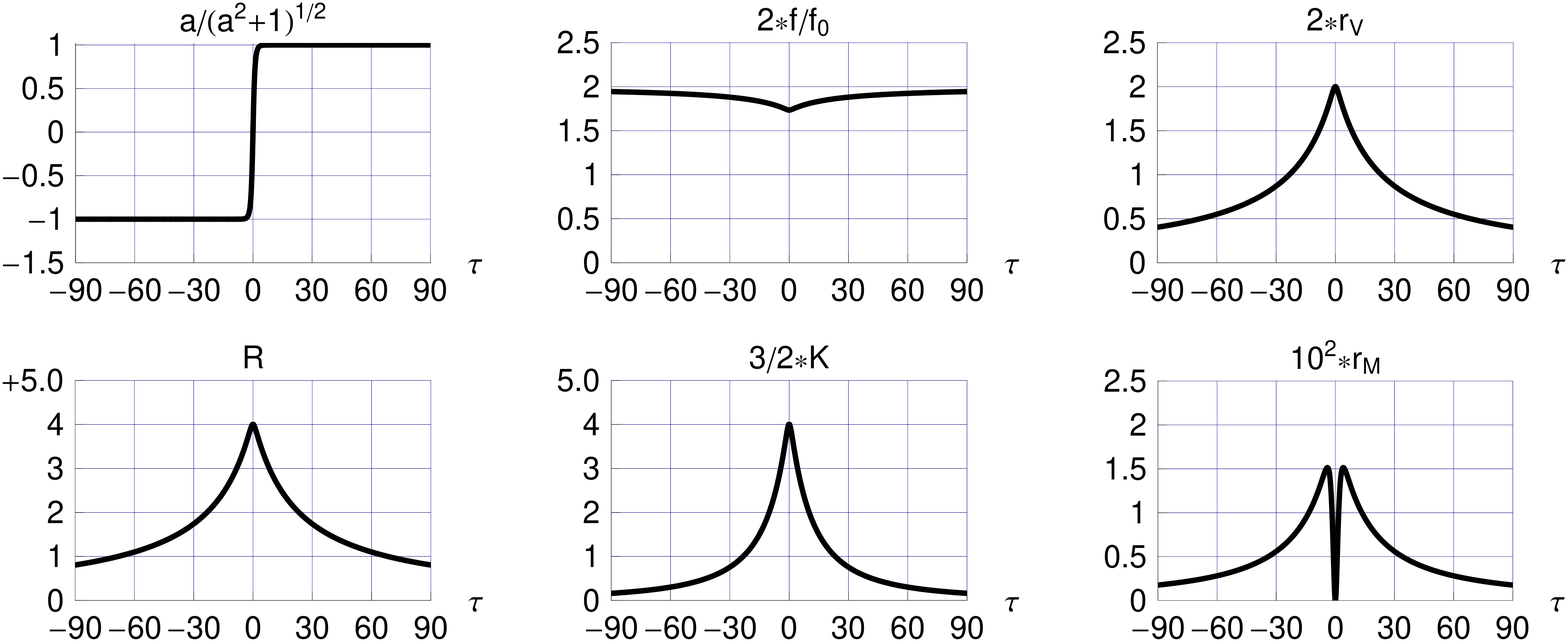}
\end{center}
\vspace*{-4mm}
\caption{Solution of the dimensionless ODEs \eqref{eq:ODEs-Milne-like}
with \textit{Ansatz} function \eqref{eq:epsilon-Ansatz-Milne-like}.
The model parameters are  
$\{w_{M},\,\lambda,\,\gamma\}=\{1/3,\,1,\,1/10\}$ and the boundary conditions
at $\tautmp=0$ are $\{a(0),\,f(0),\,r_{M}(0)\}=\{0,\, 3^{1/4},\,0\}$.
The approximate analytic solution for $\tautmp \in [-\Delta\tautmp,\,\Delta\tautmp]$
is given by \eqref{eq:zeroth-order-analytic-solution-Milne-like}
and the numerical solution is obtained for
$\tautmp \in [-\tautmp_\text{max},\,-\Delta\tautmp]
 \cup   [\Delta\tautmp,\,\tautmp_\text{max}]$
with matching boundary conditions at $\tautmp =\pm \Delta\tautmp$,
for $\Delta\tautmp=1/10$ and $\tautmp_\text{max}=4 \times 10^3$.
Shown, over a reduced time interval,
are the dynamic variables $a(\tautmp)$ and $f(\tautmp)$,
together with the corresponding
dimensionless gravitating vacuum energy density
$r_V(\tautmp)$ from \eqref{eq:rV-Ansatz-Milne-like},
the dimensionless matter energy density $r_{M}(\tautmp)$
from \eqref{eq:adot-ODE-Milne-like},
the  dimensionless Ricci curvature scalar $R(\tautmp)$  from \eqref{eq:R-from-a},
and the dimensionless Kretschmann curvature scalar $K(\tautmp)$ from \eqref{eq:K-from-a}.
The dimensionless vacuum and matter energy densities at $\tautmp=0$
are $r_{V}(0)=\lambda=1$ and $r_{M}(0)=0$.
The Minkowski equilibrium value of the dimensionless $q$ variable
is given by $f_0 \approx 1.51749$ from \eqref{eq:f0-Ansatz-Milne-like}.
}
\label{fig:Milne-like}
\vspace*{4mm}
\begin{center}
\includegraphics[width=1\textwidth]{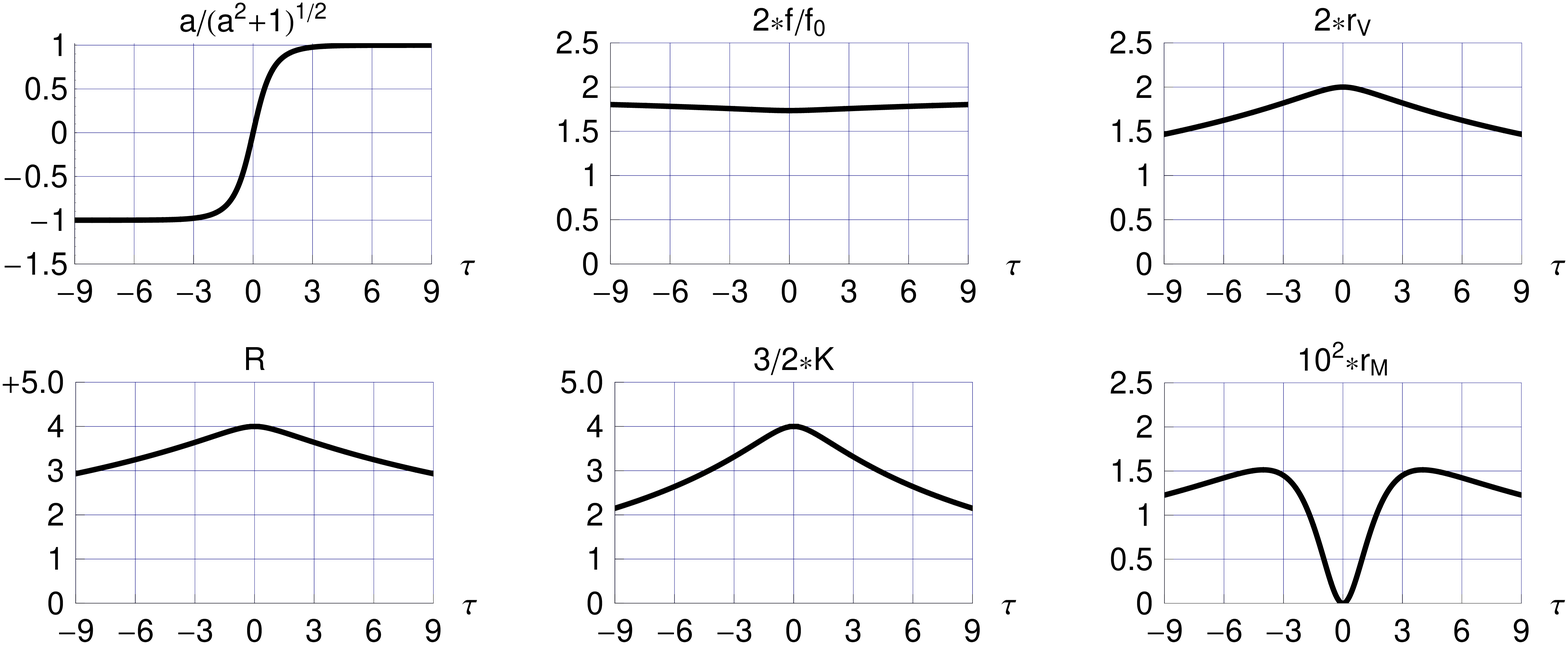}
\end{center}
\vspace*{-4mm}
\caption{Results from Fig.~\ref{fig:Milne-like} close to $\tautmp=0$.
}
\label{fig:Milne-like-NEAR}
\end{figure*}

\begin{figure*}[t]
\vspace*{-0cm}   
\begin{center}
\includegraphics[width=1\textwidth]{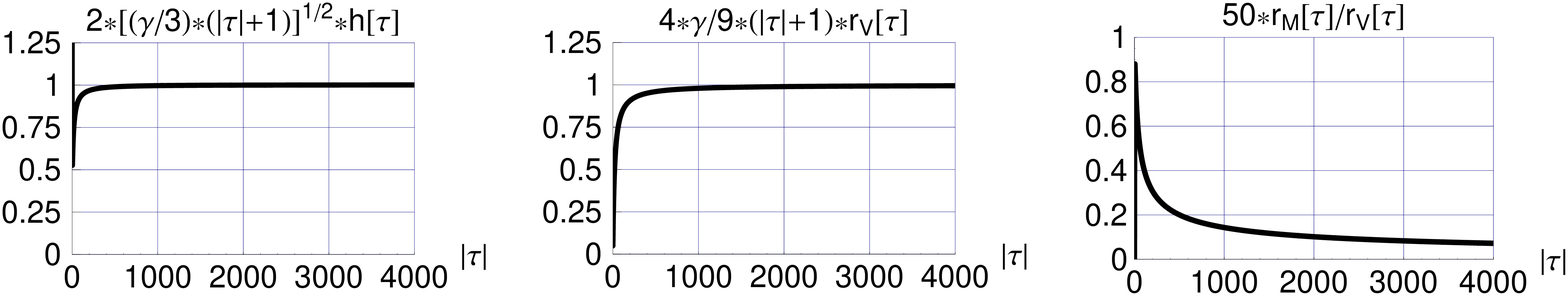}
\end{center}
\vspace*{-4mm}
\caption{Asymptotic behavior of the numerical results from Fig.~\ref{fig:Milne-like}.
}
\label{fig:Milne-like-asymptotics}
\vspace*{0mm}
\end{figure*}

\subsection{Numerical results}
\label{subsec:Numerical-results}

Numerical results are shown in Figs.~\ref{fig:Milne-like}
and \ref{fig:Milne-like-NEAR}, where the numerical procedure is
described in the caption of the first figure.
We remark that the curves of Fig.~\ref{fig:Milne-like-NEAR}, 
in particular,
are not perfectly smooth near $\tautmp=\pm \Delta\tautmp$, 
but the behavior can be improved if we add perturbative corrections
\eqref{eq:pert-solution-Milne-like} to the
zeroth-order results \eqref{eq:zeroth-order-analytic-solution-Milne-like}.
The asymptotic $|\tautmp| \to \infty$ behavior
of the numerical results is displayed in
Fig.~\ref{fig:Milne-like-asymptotics} and
agrees with the analytic behavior from \eqref{eq:asymptotic-results}.

Numerical results with nonrelativistic matter ($w_M=0$)   
and other changes are presented in 
App.~\ref{app:Crossovers-phase-transitions-density-perturbations}.

\subsection{Generalizations and attractor behavior}
\label{subsec:Generalizations}

For the boundary conditions and setup as given in
Sec.~\ref{subsec:ODEs}, we have obtained
two results: the regular behavior at $\tautmp=0$,
locally corresponding to de-Sitter spacetime,
and the attractor behavior for $|\tautmp| \to \infty$,
approaching Minkowski spacetime.
Both results hold for \emph{any} positive value of
the  cosmological constant, $\lambda>0$.
The Minkowski-spacetime attractor behavior has been
established numerically, but also follows from the analytic results
\eqref{eq:asymptotic-rV} and \eqref{eq:asymptotic-rM-over-rV},
which do not depend on the actual value of
the boundary condition $r_V(0)=\lambda>0$
(see Ref.~\cite{KlinkhamerVolovik2016}
for further discussion of this type of attractor behavior).

Similar results are obtained
for $\lambda=0$ in \eqref{eq:epsilon-Ansatz-Milne-like} and
\eqref{eq:rV-Ansatz-Milne-like}, as long as
the $\tautmp=0$ boundary conditions \eqref{eq:a-f-rM-bcs-Milne-like}
have a value $f(0) \ne \pm f_0$, so that $r_V(0) > 0$.
In fact, the general requirement for obtaining the regular behavior at
$\tautmp=0$ and the Minkowski-spacetime attractor is simply
that the boundary conditions \eqref{eq:a-f-rM-bcs-Milne-like} have
\beq\label{eq:general-requirement}
r_V(0)>0\,,
\eeq
which gives a violation of the strong energy condition  
at $\tautmp=0$ (cf. Ref.~\cite{Ling2018}).
The perturbative results from Sec.~\ref{subsec:Analytic-results}   
still hold true, as long as $\lambda$ replaced by $r_V(0)$.

The requirement \eqref{eq:general-requirement}
may even be satisfied for the case of a negative value of
the cosmological constant $\lambda$,
provided there is an appropriate value of $f(0)$, according to the $r_V$
expression \eqref{eq:rV-Ansatz-Milne-like}
for the explicit \textit{Ansatz} \eqref{eq:epsilon-Ansatz-Milne-like}.

\section{Discussion}
\label{sec:Discussion}

In this article, we have presented a physics
model for a universe which has, at $\ttmp=0$, both
a vanishing cosmic scale factor $a(\ttmp)$ and finite values of the
Ricci and Kretschmann curvature scalars, $R(\ttmp)$ and $K(\ttmp)$, 
according to the analytic results in
Sec.~\ref{subsec:Analytic-results} and the relevant panels in
Figs.~\ref{fig:Milne-like} and \ref{fig:Milne-like-NEAR}.
This regular behavior at $\ttmp=0$ differs from that of the
standard relativistic-matter Friedmann 
solution~\cite{Friedmann1922-1924,MisnerThorneWheeler2017},
which has  $a(0)=0$ and a diverging value of the Kretschmann
curvature scalar, $K(t) \propto 1/t^{4}$.
See Ref.~\cite{Klinkhamer2019} for a possible regularization
of this divergence by the introduction of a
new length scale $b$, which may result from quantum-gravity effects.  

The physics model presented in this article does not appeal,
directly or indirectly, to quantum-gravity effects.
The model relies, instead, on the standard spatially-hyperbolic ($k=-1$) 
Robertson--Walker metric \eqref{eq:RW-metric-k-is-minus-1},
as suggested in Ref.~\cite{Ling2018},
and the condensed-matter-motivated $q$-theory 
approach to the cosmological constant
problem~\cite{KlinkhamerVolovik2008a}, here
taken in the four-form-field-strength realization
\eqref{eq:classical-action}. In order to allow for
changing vacuum energy density $\rho_V(q)$, it is
possible to consider modified gravity~\cite{KlinkhamerVolovik2008b}
or to appeal to quantum-dissipative effects~\cite{KlinkhamerVolovik2016}.
The latter option is chosen in the present article,
with a particular form of vacuum-matter energy exchange.

The model spacetime constructed in this article follows from
appropriate boundary conditions at the ``origin of the universe''
[with cosmic time coordinate $\ttmp=0$ for the $k=-1$
Robertson--Walker metric \eqref{eq:RW-metric-k-is-minus-1}]:
the gravitating vacuum energy density
$\rho_V(0)$ may take any positive value, 
but the matter energy density $\rho_{M}(0)$ must be strictly zero. 
Matter creation must, in fact, take place at a moment different
from the big bang coordinate singularity
and has, in this article, been modelled
by the source term \eqref{eq:S-assumed}. 
This source term 
contains the
Zeldovich--Starobinsky $R^2(\ttmp)$ term for the
production of massless particles
by spacetime curvature~\cite{ZeldovichStarobinsky1977} and a
hypothetical on/off factor $a(\ttmp)/\sqrt{a^2(\ttmp)+1}$.
The on/off factor also makes
for time-symmetric reduced field equations and, hence, for
a time-symmetric solution.
A further comment on the dissipative nature of the source term 
\eqref{eq:S-assumed}    
is relegated to App.~\ref{app:Dissipative-vacuum-energy-decay}, 
which relies on results 
obtained in App.~\ref{app:Crossovers-phase-transitions-density-perturbations}
for a modified version of the Milne-like universe.

A related type of particle production has been considered
in Ref.~\cite{BoyleFinnTurok2018}, with ``particles'' appearing
to comoving observers in the post-big-bang phase and
``antiparticles'' to comoving observers in the pre-big-bang phase
(see also the discussion of the last paragraph in   
App.~\ref{app:Dissipative-vacuum-energy-decay}).
The idea of a CPT-invariant vacuum state and a
CPT-symmetric universe (or rather an universe--anti-universe pair)
is certainly interesting,
but the metric used in Ref.~\cite{BoyleFinnTurok2018} is
singular and not derived from the Einstein equation
(or an improved version thereof).
The hope is that the time-symmetric spacetime  manifold of our model,
derived from the Einstein equation and the generalized Maxwell equation
for the $q$-field, may provide the proper setting for   
a CPT-symmetric universe, if at all relevant.

\begin{acknowledgments}

F.R.K. thanks G.E. Volovik for discussions on vacuum energy decay.

\end{acknowledgments}

\begin{appendix}

\section{Crossovers, phase transitions, and density perturbations}
\label{app:Crossovers-phase-transitions-density-perturbations}

In this appendix, we present a modified version of the Milne-like universe of 
Sec.~\ref{sec:Model-for-a-Milne-like-universe}. The main changes
are twofold. First, we introduce nonrelativistic matter
to be used as a diagnostic tool
(in principle, it is also possible to have additional relativistic matter) 
and, second, we consider possible
effects from crossovers and phase transitions (giving
rise to, respectively, a change of the coupling parameter 
controlling vacuum-matter energy exchange and the creation of
matter density perturbations).

The goal of these changes is to obtain Friedmann-like
phases, where the matter energy density $\rho_M(t)$ is dominant
and where there is  more or less the standard Hubble 
expansion, $|a(t)| \propto |t|^n$ with $n \approx 2/3$.
In these Friedmann-like phases of the modified Milne-like universe, 
perturbations of the matter density
may be created by further phase transitions (for example,
by the collisions of bubbles in a first-order phase transition)
and we establish the growth of these perturbations.
This last result is an essential input for the discussion
of App.~\ref{app:Dissipative-vacuum-energy-decay}.

The dynamics of the modified Milne-like universe is
governed by the ODEs \eqref{eq:ODEs-Milne-like}  
with $w_M=0$ (nonrelativistic matter)
and the coupling constant $\gamma$ in 
\eqref{eq:fdot-ODE-Milne-like}  
replaced by the following time-dependent coupling parameter:
\beq
\label{eq:gamma-modified}
\gamma(t)= \gamma_\text{bb} + \frac{t^6}{t^6+t_\text{cross}^6}\,\Delta\gamma\,,
\eeq
where $t_\text{cross}>0$ is the crossover time scale.

\begin{figure*}[t] 
\vspace*{-0mm}
\begin{center}
\includegraphics[width=1\textwidth]{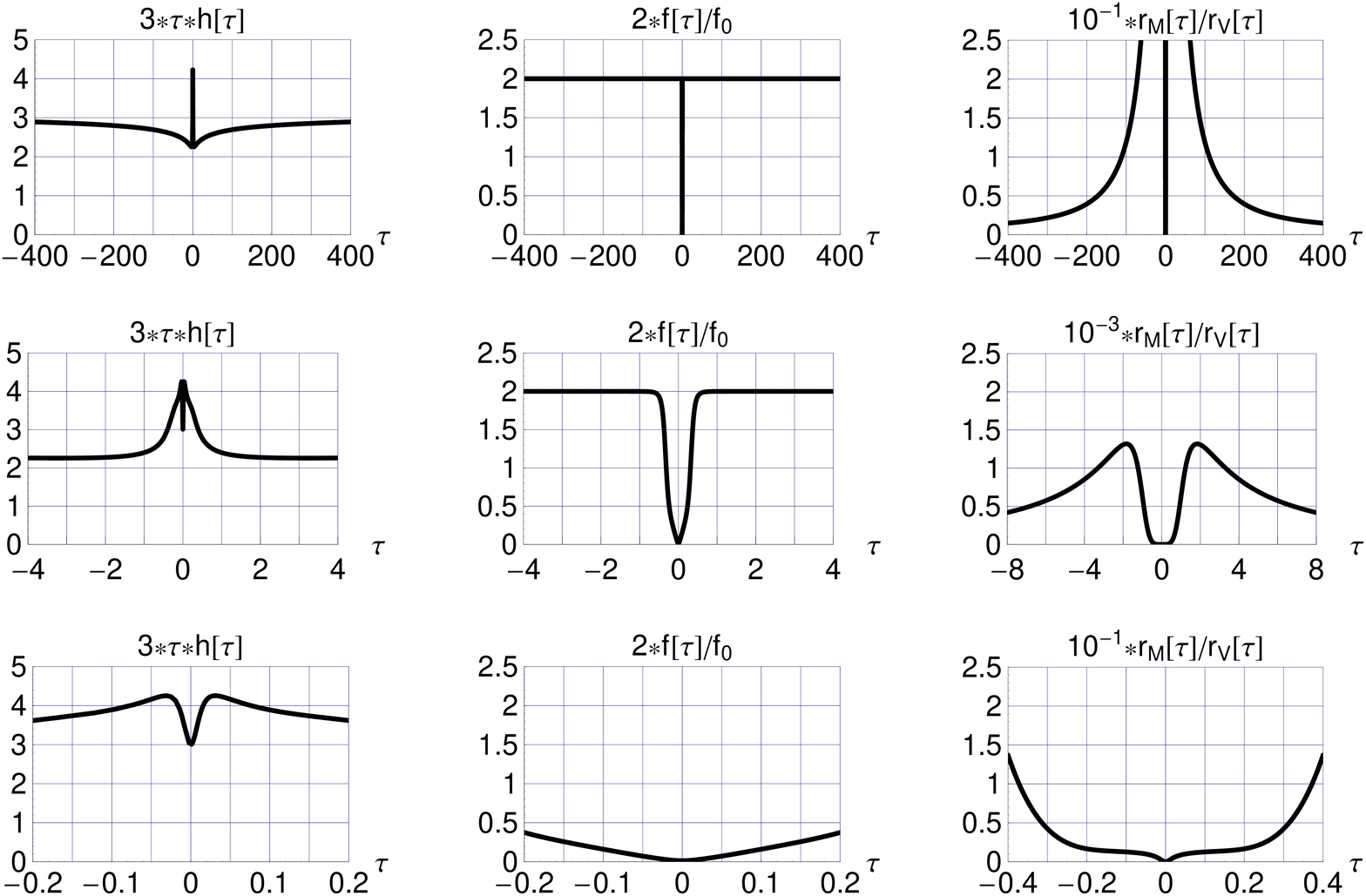}
\end{center}
\vspace*{-4mm}
\caption{Same as Fig.~\ref{fig:Milne-like} but now with
model parameters $\{w_{M},\,\lambda\}=\{0,\,1\}$ and
vacuum-matter-energy-exchange coupling $\gamma(\tau)$ from
\eqref{eq:gamma-modified} with parameter values
$\{\gamma_\text{bb},\, \Delta\gamma,\, \tau_\text{cross}\}
= \{1,\, 10^{4},\,1\}$. The boundary conditions
at $\tautmp=0$ are $\{a(0),\,f(0),\,r_{M}(0)\}=\{0,\, 1/100,\,0\}$.
The numerical calculation uses $\Delta\tau=10^{-3}$.
The dimensionless vacuum and matter energy densities at $\tautmp=0$
are $r_{V}(0)\approx 3\times 10^{4}$ and $r_{M}(0)=0$.
For the model parameters and boundary conditions chosen, the ratio
of $r_{M}/r_{V}$ reaches a peak value of order $10^{3}$
at $\tau\approx \pm 2$ and drops to unity at $\tau\approx \pm 600$.
The top-right panel shows the behavior of the
ratio $r_{M}/r_{V}$ for $|\tau| \gg 1$
and the bottom-right panel the behavior close 
to the big bang coordinate singularity at $\tau=0$.
The middle-row panels show that there are two phases
around $\tau= \pm 3$, where the matter energy density dominates
over the vacuum energy density and the expansion is Friedmann-like
\eqref{eq:Friedmann-like-expansion}.
}
\label{fig:modified-Milne-like}
\vspace*{0mm}
\end{figure*}

Numerical results are presented in Fig.~\ref{fig:modified-Milne-like} 
with boundary conditions at $\tau=0$ giving $r_V(0) \gg \lambda >0$
and appropriate parameter values for the 
vacuum-matter-energy-exchange coupling \eqref{eq:gamma-modified}.
Recall that $\tau$ is the dimensionless cosmic time coordinate introduced 
in Sec.~\ref{subsec:ODEs}, specifically, 
\beq
\label{eq:tau-def}
\tau\equiv t\,c/l_\text{planck}\,,
\eeq
in terms of the Planck length \eqref{eq:l-planck}.
[The cosmic scale factor $a(\tau)$ is not shown explicitly
in Fig.~\ref{fig:modified-Milne-like}, but its behavior is
similar to that in Figs.~\ref{fig:Milne-like} and 
\ref{fig:Milne-like-NEAR}.]
The model universe of Fig.~\ref{fig:modified-Milne-like} 
has two phases near $\tau=\pm 3$ where the expansion is Friedmann-like,
\bsubeqs\label{eq:Friedmann-like-expansion}
\beqa
\label{eq:Friedmann-like-expansion-a}
a(t) &\propto& \sgn(t)\,|t|^n\,,
\\[2mm]
\label{eq:Friedmann-like-expansion-H}
H(t) &\sim& n\,t^{-1}\,,
\\[2mm]
\label{eq:Friedmann-like-expansion-n}
n &\approx& 3/4\,.
\eeqa
\esubeqs
We remark that it is possible to reduce the large value   
of $r_{V}(0)$ in the results of Fig.~\ref{fig:modified-Milne-like} 
by choosing somewhat different model parameters and boundary conditions.
With  $\{w_M,\, \lambda\}= \{0,\, 1/10\}$,
$\{\gamma_\text{bb},\, \Delta\gamma,\, \tau_\text{cross}\}= \{1,\, 10^{2},\,1\}$, 
and $\{a(0),\,f(0),\,r_{M}(0)\}$ $=$ $\{0,\, 1/\sqrt{2},\,0\}$,
for example, we get $r_{V}(0) = 5.6$ and
$n \approx 0.96 < 1$ for the Hubble 
parameter \eqref{eq:Friedmann-like-expansion-H}
at $\tau \approx 4$.

It is also possible to imagine that first-order phase transitions, 
just after the crossovers at $t=\pm t_\text{cross}$ 
relevant to \eqref{eq:gamma-modified},
generate perturbations of the matter density.
We refer to Chaps.~6 and 7 in Ref.~\cite{Mukhanov2005}
for a general discussion of the dynamic evolution of these
density perturbations and the corresponding scalar
metric perturbations.

Consider the Fourier transform $\delta_\mathbf{k}(t)$ of the
relative matter density perturbation, defined by
\beq
\label{eq:Fourier-transformation-density-perturbation}
\frac{\delta\rho_M(t,\,\mathbf{q})}{\overline{\rho}_M(t)}
\equiv
\frac{\rho_M(t,\,\mathbf{q})-\overline{\rho}_M(t)}{\overline{\rho}_M(t)}
= \int \frac{d^3 k}{(2\pi)^3}\,
\delta_\mathbf{k}(t)\, \exp\big[i\,\mathbf{k}\cdot\mathbf{q}\big]\,,
\eeq
where $\mathbf{q}$ are the Lagrangian spatial coordinates, which are 
comoving with respect to the Hubble flow~\cite{Mukhanov2005}.
Then, the Newtonian analysis of subhorizon adiabatic density
perturbations in an expanding homogeneous and isotropic universe 
gives the following ODE 
(the overdot stands for differentiation with respect to
$\ttmp$):
\beq\label{eq:delta-ODE}
\ddot{\delta}_\mathbf{k}(t) + 2\,H(t)\,\dot{\delta}_\mathbf{k}(t) 
+\left(\frac{c_s^2\,|\mathbf{k}|^2}{a^2(t)}
        - 4 \pi G_N\,\overline{\rho}_M(t)\right) 
        \delta_\mathbf{k}(t)=0\,,
\eeq
which corresponds to Eq.~(6.47) in Ref.~\cite{Mukhanov2005}. 
The $c_s^2\,|\mathbf{k}|^2/a^2$ term in \eqref{eq:delta-ODE}
can be neglected for long-wavelength (but still subhorizon) perturbations.
With arbitrary time-independent functions $C_1(\mathbf{k})$ 
and $C_2(\mathbf{k})$, 
the general long-wavelength solution of
\eqref{eq:delta-ODE} without $c_s^2\,|\mathbf{k}|^2/a^2$ term is given by
\beq
\label{eq:delta-ODE-general-solution}
\delta_\mathbf{k}(t)
=
C_1(\mathbf{k}) \,H(t)\,\int^{t}\,\frac{dt'}{a^2(t')\,H^2(t')}
+ C_2(\mathbf{k})\,H(t)\,,
\eeq
which corresponds to Eq.~(6.55) in Ref.~\cite{Mukhanov2005}.

In the context of our bouncing cosmology with an odd
Hubble function $H(t)$,
we now make the following crucial observation:
\emph{the decreasing $C_{2}$ term of \eqref{eq:delta-ODE-general-solution} 
is odd in $t$, whereas the increasing $C_{1}$ term 
of \eqref{eq:delta-ODE-general-solution} is even in $t$.}
A trivial check of this statement
can be performed with the functions $a(t) = t$ and $H(t) = 1/t$.

If we now consider the possibility that phase transitions at
$t = \pm t_\text{trans}$, for $t_\text{trans} \sim t_\text{cross} >0$,
generate matter density perturbations 
\eqref{eq:Fourier-transformation-density-perturbation},
then the generic matter density perturbations (possibly different
at $t = \pm t_\text{trans}$) grow with $t \rightarrow \infty$
for $t > t_\text{trans}$
and grow with $|t| \rightarrow \infty$ for $t < -t_\text{trans}$.

Specifically, we have, in the model universe of Fig.~\ref{fig:modified-Milne-like}, 
two Friedmann-like phases \eqref{eq:Friedmann-like-expansion}
around dimensionless cosmic time coordinates $\tau = \pm 3$.
In these two phases, we get from \eqref{eq:Friedmann-like-expansion}
and \eqref{eq:delta-ODE-general-solution}
\bsubeqs
\beqa\label{eq:delta-ODE-specific-solution}
\delta_\mathbf{k}(t)&=&
%
\begin{cases}
 \delta_\mathbf{k}^{+}\,\Big(t^2/t_\text{trans}^2\Big)^{1-n} 
 + \delta_\mathbf{k}^{-}\;t_\text{trans}/t\,,   
 &  \text{for}\quad t_\text{trans} \leq |t| < t_\text{approx}\,,
 \\[2mm]
 0 \,,   
 &  \text{for}\quad |t| < t_\text{trans} \,,
\end{cases}
\\[2mm]
\label{eq:delta-ODE-specific-solution-deltapm-def}
\delta_\mathbf{k}^{\pm} &\equiv&
\frac{1}{2}\,\Big[
 \delta_\mathbf{k}\big(t_\text{trans}\big)
 \pm  \delta_\mathbf{k}\big(-t_\text{trans}\big)
\Big]\,,
\eeqa
\esubeqs
where the numerical value \eqref{eq:Friedmann-like-expansion-n} is assumed 
to be a reasonable approximation over the time interval 
$(-t_\text{approx},\,-t_\text{trans}]$ $\cup$ $[t_\text{trans},\,t_\text{approx})$.
The power $p\equiv 1-n$ of the growing mode 
in \eqref{eq:delta-ODE-specific-solution} takes the approximate numerical 
value $p \approx 1/4$ for the matter-dominated phases 
\eqref{eq:Friedmann-like-expansion} around  $\tau = \pm 3$
in Fig.~\ref{fig:modified-Milne-like}. As mentioned before,
the result \eqref{eq:delta-ODE-specific-solution}
will be used in App.~\ref{app:Dissipative-vacuum-energy-decay}.

\section{Dissipative vacuum energy decay}
\label{app:Dissipative-vacuum-energy-decay}

In this appendix, we comment on the interpretation of the
source term \eqref{eq:S-assumed}. 
The $a(t)$ solution obtained in Secs.~\ref{subsec:Analytic-results}
and \ref{subsec:Numerical-results}
(see also Figs.~\ref{fig:Milne-like} and \ref{fig:Milne-like-NEAR}) 
has the following structure:
\begin{itemize}
  \item 
the big bang coordinate singularity at $t = 0$ 
with cosmic scale factor $a(0) = 0$,
  \item 
the post-big-bang phase for $t > 0$ with
$a(t) > 0$ and Hubble expansion $H(t) > 0$,
  \item 
  the
pre-big-bang phase for $t < 0$ with $a(t) < 0$ 
and Hubble contraction $H(t) < 0$.
\end{itemize}
The compound adjectives ``pre-big-bang'' and ``post-big-bang''
in the above items refer only to the cosmic time coordinate $t$.

Now, it is important to realize that $t$ is just a coordinate
and that the physically relevant time may appear to run
differently. In fact, the results from 
App.~\ref{app:Crossovers-phase-transitions-density-perturbations},
in particular the expression \eqref{eq:delta-ODE-specific-solution},
show that the modified Milne-like universe has two phases where
matter density perturbations grow with increasing $|t|$.
This implies that the corresponding thermodynamic
arrow of time (increasing entropy)
runs parallel to the cosmic time coordinate $t$
in the post-big-bang phase and opposite to $t$ in the
pre-big-bang phase. We can, then, define the following
thermodynamic time coordinate $\mathcal{T}$:
\beq
\label{eq:thermodynamic-time-coordinate}
\mathcal{T}
\equiv
\begin{cases}
 t \,,   &   \text{for}\quad t>0 \,,
 \\[2mm]
 -t \,,   &  \text{for}\quad t\leq 0\,.
\end{cases}
\eeq
A result similar to \eqref{eq:thermodynamic-time-coordinate} 
was obtained in Ref.~\cite{BoyleFinnTurok2018}, but the metric
used there has a big bang curvature singularity, which
makes the discussion of Ref.~\cite{BoyleFinnTurok2018} somewhat formal.

With the odd behavior of $a(t)$ and the new time coordinate $\mathcal{T}$
from \eqref{eq:thermodynamic-time-coordinate},
the vacuum-energy evolution equation 
\eqref{eq:rhoVdoteq} with source term \eqref{eq:S-assumed} becomes
\beq
\label{eq:rhoVdoteq-dissipative}
\frac{d \rho_V(q)}{d \mathcal{T}} =- |\mathcal{S}(\mathcal{T})|\,,
\eeq
for the homogeneous $q$-field, $q=q(\mathcal{T})$.
The evolution equation \eqref{eq:rhoVdoteq-dissipative}
corresponds to a \emph{dissipative}
process, where the ordered vacuum energy is transferred to 
the energy of disordered
particles as the thermodynamic time $\mathcal{T}$ increases. 

This notion of dissipation agrees with  
the physical picture suggested in Ref.~\cite{BoyleFinnTurok2018},  
namely that the pre-big-bang and post-big-bang phases 
mentioned in the first paragraph of this appendix
correspond to an ``anti-universe'' and a ``universe,'' respectively, 
both created from the vacuum at $\mathcal{T}=0$ and evolving forward
with time $\mathcal{T}>0$.

\end{appendix}

\end{document}